\documentclass[twocolumn,preprintnumbers,amsmath,amssymb]{revtex4}

\usepackage{graphicx}% Include figure files
\usepackage{dcolumn}% Align table columns on decimal point
\usepackage{bm}% bold math

\begin{document}

\title{Fiber-assisted detection with photon number resolution}

\author{Daryl Achilles, Christine Silberhorn, Cezary \'Sliwa, Konrad Banaszek, and Ian A. Walmsley}

\affiliation{Clarendon Laboratory, University of Oxford, Parks
Road, Oxford OX1 3PU, UK}

\date{\today}

\begin{abstract}

We report the development of a photon-number resolving detector
based on a fiber-optical setup and a pair of standard avalanche
photodiodes. The detector is capable of resolving individual
photon numbers, and operates on the well-known principle by which
a single mode input state is split into a large number (eight) of
output modes. We reconstruct the photon statistics of weak
coherent input light from experimental data, and show that there
is a high probability of inferring the input photon number from a
measurement of the number of detection events on a single run.
\end{abstract}

\maketitle

Many quantum information strategies require the preparation of
nonclassical states. For example, the method of linear optical
quantum computing proposed by Knill, Laflamme, and
Milburn~\cite{KLM} demands the preparation of single photon states
as well as maximally entangled photon multiplets.  A number of
schemes have been proposed for the preparation of such states,
including single photon emitters~\cite{singleph, YamamotoSP,
Rempe} and conditionally prepared photon pairs from parametric
downconversion~\cite{Konrad,Pittman,Loock}.  Conditional state
preparation requires the ability to distinguish states of
different photon number, which is not possible using conventional
photodetectors. Photon number resolution is also desirable to
enhance the security of quantum cryptographic
schemes~\cite{Hwang,Christine}. In this case it is important to
measure the photon statistics of the source at the sending and
receiving stations. For implementations that use weak coherent
states this means distinguishing between the detection of, say,
one or two photons.

According to the quantum theory of photodetection, the signal
obtained from an ideal noise-free detector has a discrete form
corresponding to the absorption of an integer number of quanta
from the incident radiation. In practice, however, the granularity
of the output signal is concealed by the noise of the detection
mechanism. When very low light levels are detected using devices
with single-photon sensitivity such as photomultipliers or
Geiger-mode operated avalanche photodiodes (APDs), the electronic
signal can be reliably converted into a binary message telling us
with high efficiency whether an absorption event has occurred or
not. However, the intrinsic noise of the gain mechanism necessary
to bring the initial energy of absorbed radiation to the
macroscopic level completely masks the information on exactly how
many photons have triggered that event.

There are several methods for constructing photon number resolving
detectors.  Among those demonstrated to date are the segmented
photomultiplier~\cite{yamamoto}, the superconducting
bolometer~\cite{NIST}, and the superconducting transimpedance
amplifier~\cite{sobolewski}.  These detectors operate at cryogenic
temperatures and have single photon quantum efficiencies ranging
from about 20\% for the superconducting devices to approximately
70\% in the case of the segmented photomultiplier. On the other
hand, conventional room temperature APDs have intrinsic quantum
efficiencies up to 80\%, though they respond only to the presence
or absence of radiation. The ease of use of these devices suggest
that it is worth exploring ways to develop photon resolving
capability.

In this paper, we present an experimental detection scheme that
implements a full photon-number resolved measurement of light
intensity for optical pulses. Our experimental setup  follows the
main idea outlined in~\cite{BanaszekWalmsley}.  The detection is
done by splitting the input pulse into separate parts that are
expected to contain no more than one photon and then detecting
them with conventional avalanche photodiodes. The setup we
describe here is constructed only from standard passive fiber
optic components, but nevertheless permits the splitting into
arbitrary many time-delayed pulses, while retaining a fixed number
of output spatial modes. The design is depicted schematically in
Fig.~1. The basic elements are 50/50 couplers and fibers of
variable lengths. In the first step the input signal is divided
into two and launched into two fibers of unequal lengths, such
that the partite pulses are delayed with respect to each other.
Their subsequent combination at another 50/50 coupler then yields
two pulses in each of the two channels. A second stage in which
the relative delay is twice as long as the previous stage provides
four pulses per output channel. Further doubling of the number of
temporal output modes can be achieved by adding more stages. Thus
the photon number of the incident pulse can then be detected using
only two APDs, if the time separation is ensured to exceed the
dead times of the APDs. A less efficient scheme using a single
coupler and a fiber loop has also been investigated~\cite{czech}.

\begin{figure}\centerline{\scalebox{0.40}{\includegraphics{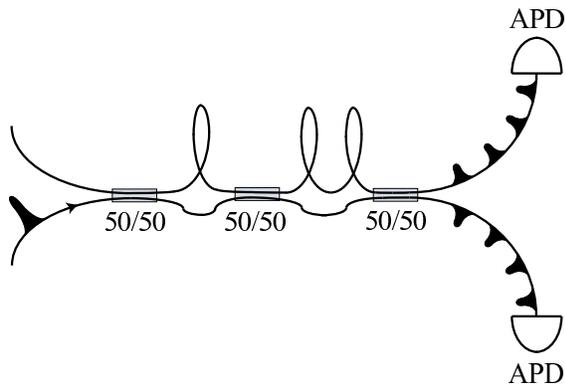}}}
\caption{Schematic setup of the detector; 50/50: symmetric fiber
couplers, APD: avalanche photodiode.} \label{fig:scheme}
\end{figure}

For our implementation we used single mode fibers at 780nm (Lucent
SMC-AO780B) to construct a detector with two stages,  i.e. eight
temporal modes. After the fiber setup the pulses were separated in
time by 108~ns to 164~ns.  Without insertion losses  the
transmission through the complete fiber system  was measured to be
higher than 56\%. A standard laser diode (Thorlabs V3-780-TO-DA)
driven in pulsed operation at 777~nm served as a source of
coherent states. These pulses had a width of approximately 14~ns
and a repetition rate of 10~kHz.  The detection of the attenuated
signals at the  fiber outputs was carried out with two standard
APDs~(Perkin Elmer SPCM-AQR-13-FC), which are specified to have an
efficiency of 66\% and typical deadtimes of 50~ns. Hence the time
bins with non-zero photons could be identified using a digital
oscilloscope.

A careful analysis of the measured data is crucial since the
statistics can emulate ``non-classical" properties such as
sub-Poissonian photon number distributions. This is mainly due to
the fact that high photon numbers result in a non-negligible
probability that two photons remain together in one pulse and are
counted as one. Therefore the number of stages of the fiber
configuration essentially limits the incident photon numbers that
can be reliably distinguished by the detector. The probability
$p(k)$ of the detected counts is linked to the signal
photon-number distribution $\rho(n)$ by
\begin{equation}
p(k) = \sum_{n} p_{kn}(k|n) \rho(n). \label{Eq:prob}
\end{equation}
The photon number distribution can be reconstructed, if the
conditional probabilities $p_{kn}(k|n)$ of obtaining $k$ counts
for $n$ incident photons are known. The exact values of
$p_{kn}(k|n)$ depend on the detailed experimental setup including
non-ideal 50/50 splitting and unbalanced losses. For known
parameters of the fiber system they can be calculated by a basic
stochastic model that takes into account the different
possibilities for spreading the incident photons into the output
modes. The inversion of the associated matrix allows the
identification of the incident photon number distribution $
\rho(n)$ from the measured distribution $p(k)$. For our detector
with eight output pulses we only expected a good performance  for
distributions with mean photon numbers smaller than three. Hence
we could restrict our data processing to $p_{kn}(k|n)$ with $n$
smaller than eight, which gives an estimated error being smaller
than one percent.

\begin{figure}\centerline{\scalebox{0.45}{\includegraphics{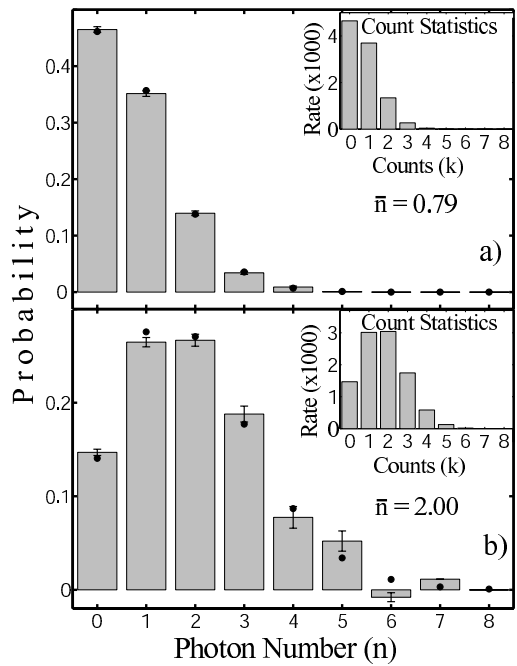}}}
\caption{\label{fig:data}Photon number statistics reconstructed by
direct inversion of the measured data (see insets) from the fiber
based detector. The data (bars) shows good agreement with the
theoretical fits (dots) of a Poissonian distribution with the same
mean photon number. }
\end{figure}

\begin{table*}
\caption{\label{tab:table1} Conditional probabilities
$\tilde{p}_{nk}(n=l|k=l)$, that $l$ counts are  triggered by  $l$
photons for Poissonian input statistics with average photon number
$\bar n$ and different losses $ 1- \eta $.}
 \begin{ruledtabular}
\begin{tabular}{|c|cccc|ccc|ccc|}
{} & \multicolumn{4}{c|}{$\eta=1$} &
\multicolumn{3}{c|}{$\eta=0.7$}& \multicolumn{3}{c|}{$\eta=0.5$}
\\
\hline $l$ & $\bar n = 0.25$& $\bar n = 0.5$ & $\bar n=1.0$ &
$\bar n=1.5$ & $\bar n = 0.25$ & $\bar n = 0.5$ & $\bar n=1.0$ &
$\bar n = 0.25$ & $\bar n=0.5$ & $\bar n = 1.0$ \\
\hline
1&0.984&0.969&0.938&0.908&0.918&0.842&0.709&0.876&0.767&0.588\\
2&0.969&0.939&0.881&0.825&0.908&0.824&0.678&0.869&0.755&0.569\\
3&0.954&0.910&0.826&0.750&0.898&0.806&0.649&0.862&0.743&0.552\\
\end{tabular} \end{ruledtabular}
\end{table*}

To simplify the inversion the influence of the convolution and the
losses can be considered separately.  This is equivalent to
assuming that the probability of sending an input photon to all
the output channels sums to one and introducing an overall loss
factor.  Our reconstructions of the photon number from the
experimental data do not correct for losses, since these leave the
form of the Poissonian distributions unchanged and reduces the
means only. In addition, by not including the losses we
demonstrate the potential capability of the detector to
distinguish between different photon numbers if the transmission
through the fibers is optimized and highly efficient detectors are
available.  Fig.~2 shows our experimental results for coherent
states with mean photon number of 0.79 and 2.00 respectively. The
insets depict the actual count statistics we directly obtained
from the two APDs for samples of $10^4$ single measurements. To
eliminate dark counts and afterpulsing we applied a temporal
gating such that only counts within the expected time windows of
$45$ ns were accepted. The bars in the main graphs represent the
normalized photon number distributions we acquired from the
detected data using the inversion of $p_{kn}(k|n)$. We determined
the given error bars by running 1000 Monte Carlo simulations  for
our measurement samples. The dots indicate the theoretical fits
for the respective distributions with the same average photon
numbers as the experimental data. The analysis reveals excellent
agreement between experiment and theory for distributions with
average photon numbers smaller or equal to one (see Fig.~2a).
Moreover, as seen in Fig.~2b, the direct inversion still works
reasonably well for experimental data corresponding to a coherent
state with mean photon number of $2.00$, though in this case
negative probabilities may appear. Note, however, that we did not
optimize the inversion in any way nor did we put any constraints
that would exclude negative probabilities.

We tested our inversion for increasing mean photon numbers with
simulated Monte Carlo data for input Poissonian statistics, which
we truncated at a photon number of eight. In this way we were able
to confirm that direct inversion is only possible up to a mean
photon number two. The error in these reconstructions, based on
the mean square differences between the distributions, is
$4.17\times10^{-4}$.  The comparison of the experimental data with
simulated data revealed that the inversion is very sensitive to
events that measure counts with $k>6$.  These events cause
negative probabilities to emerge in the photon distribution due to
the inversion. Using advanced estimation techniques that exploit
known probability properties should allow accurate reconstruction
of distributions with higher mean photon
number~\cite{KonradEstimation}.  Following that line we
investigated coherent states with averages up to four. If we
assume Poissonian distributions, we can estimate the mean of the
detected data by a simple maximum likelihood estimation, taking
into account the conditional matrix $p_{kn}(k|n)$. Using this
approach we found good consistency between experimental data and
the corresponding theoretical distribution for a mean photon
number of $3.78$, where the deviations for all possible photon
numbers summed up to be smaller than $0.012$. Thus the photon
resolving detector proves to be a valuable tool to explore quantum
states in the photon number basis.

Generally speaking there exist two main tasks where photon number
resolving detectors are needed. So, far we have discussed one case
when the initial photon distribution is either unknown or---like
in quantum cryptography---is to be confirmed by ensemble
measurements.  In the second case the incident state is well known
from the beginning, and one wants to perform the detection of the
photon number on a single shot basis to address states
individually. In such applications, e.g. conditional state
preparation~\cite{Konrad}, the performance of the detector has to
be evaluated on the confidence that $l$ counts have actually been
triggered by $l$ photons. In our experiment this confidence can be
characterized by the conditional probabilities
$\tilde{p}_{nk}(n=l|k=l)$, which describes the probability that
$l$ photons cause $l$ detection events. The
$\tilde{p}_{nk}(n=l|k=l)$ always depend on  the losses of the
complete system as well as on the photon statistics, i.e. on the
input state itself. To illustrate the reliability of our detector
in such settings we have calculated the relevant conditional
probabilities for input Poissonian photon statistics and different
losses~(see Table~\ref{tab:table1}). We find that a lossless
detector with eight temporal modes allows appropriate
discriminations of low photon numbers for coherent states with
mean photon numbers up to $\bar{n}=1.5$. The distribution of the
input light into a limited number of modes can therefore be
tolerated if the ratio of maximum photon number to outgoing modes
is sufficiently small. As expected losses decrease the confidence
for measurements on single quantum systems, which restricts the
usage of the detector to photon number distribution with low
enough mean values.

In summary we have demonstrated a fiber based photon detector,
which is capable of resolving multiple photons. In our particular
setup with eight temporal modes we could measure and recover the
photon number statistics by direct inversion for coherent states
with  mean photon numbers up to two. The presented design for the
photon-counting detector can  easily be extended to further split
the incident pulse in time without the need to increase the number
of  spatial modes, which are monitored by the two APDs. Adding on
stages allows both an increase in the reliability of the detector
and permits the resolution of higher photon numbers. Because of
the comparative simplicity of the setup it is feasible to
establish this photon-counting detector as a standard device in
quantum optics. In the context of quantum information processing
this is a step forward in implementing more sophisticated schemes
based on conditional state preparation.

We are grateful for conversations with Czes{\l}aw Radzewicz. This
research was supported by the US Department of Defense through the
Army Research Office via grant number DAAD19-02-1-0163.  C.
{\'S}liwa's current address is:  Centre for Theoretical Physics,
Al. Lotnikow 32/46, 02-668 Warszawa, Poland.  After completion of
this experiment, we became aware of very similar work performed
independently~\cite{Fitch}.


\begin{thebibliography}{10}
\bibitem{KLM} E. Knill, R. Laflamme, and G. J. Milburn, Nature {\bf 409}, 46
(2001).

\bibitem{singleph} P. Michler, A. Kiraz, C. Becher, W. V. Schoenfeld,
P. M. Petroff, L. Zhang, E. Hu, and A. Imamoglu, Science {\bf
290}, 2282 (2000).

\bibitem{YamamotoSP} C. Santori, M. Pelton, G. Solomon, Y. Dale, and Y. Yamamoto,
Phys. Rev. Lett. {\bf 86}, 1502 (2001).

\bibitem{Rempe} A. Kuhn, M. Hennrich and G. Rempe, Phys. Rev. Lett. {\bf 89}, 067901 (2002).

\bibitem{Konrad} C. {\'S}liwa and K. Banaszek, Phys. Rev. A {\bf 67}, 030101(R) (2003).

\bibitem{Pittman} T.B. Pittman, M.M. Donegan, M.J. Fitch, B.C. Jacobs, J.D.
Franson, P. Kok, H. Lee, and J.P. Dowling, quant-ph/0303113
(2003).

\bibitem{Loock} P. van Loock and N. L\"utkenhaus, quant-ph/0304057 (2003).

\bibitem{Hwang} W. Y. Hwang, quant-ph/0211153 (2003).

\bibitem{Christine}J. Calsamiglia, S. M. Barnett, and N.
Luetkenhaus, Phys. Rev. A {\bf 65}, 012312 (2002).

\bibitem{yamamoto} J. Kim, S. Takeuchi, Y. Yamamoto, and H. H.
Hogue, Appl. Phys. Lett.  {\bf 74},  902 (1999).

\bibitem{NIST}B. Cabrera, R. M. Clarke, P. Colling, A. J. Miller,
S. Nam, and R. W. Romani, Appl. Phys. Lett. {\bf 73}, 735 (1998).

%S. W. Nam, D. Rosenberg, A.J. Miller, A. Salinen, E.
%Grossman, and J. M. Martinis, ``Photon-Number Resolving Detectors
%for Quantum Communications,'' presented at the NIST Workshop on
%Single-Photon: Detectors, Applications, and Measurement Methods,
%Gaithersburg, M.D., 31 March - 1 April 2003.

\bibitem{sobolewski} S. Somani, S. Kasapi. K. Wilsher, W. Lo,
R. Sobolewski and G. N. Gol'tsman, J. Vac. Sci. Technol. B {\bf
19}, 2766 (2001).

\bibitem{BanaszekWalmsley} K. Banaszek and I. A. Walmsley, Opt. Lett.
{\bf 28}, 52 (2003).

\bibitem{czech}  O. Haderka, M. Hamar, and J. Perina Jr.,  quant-ph/0302154  (2003).

\bibitem{KonradEstimation} K. Banaszek, Phys. Rev. A {\bf 57}, 5013 (1998).

\bibitem{Fitch} M. J. Fitch, B. C. Jacobs, T. B. Pittman, and J. D. Franson,
Applied Physics Laboratory,  Submitted for Publication,
quant-ph/0305193 (2003).

\end{thebibliography}
\end{document}